\documentclass[onecolumn]{emulateapj}
\usepackage{lscape}
\usepackage{epsfig}
\submitted{{\it Accepted for publication in AJ}}
\shortauthors{Davenport et al.}
\shorttitle{Photometric Calibrations for Red Stars with the SDSS Photometric Telescope}

\begin{document}

\title{Improved Photometric Calibrations for Red Stars Observed with the SDSS Photometric Telescope}

\author{James R. A. Davenport\altaffilmark{1,2},
John J. Bochanski\altaffilmark{2},
Kevin R. Covey\altaffilmark{3},
Suzanne L. Hawley\altaffilmark{2}, 
Andrew A. West\altaffilmark{4},
Donald P. Schneider\altaffilmark{5}}

\altaffiltext{1}{Corresponding author: uwjim@astro.washington.edu}
\altaffiltext{2}{Department of Astronomy, University of Washington, Box 351580, Seattle, WA 98195}  
\altaffiltext{3}{Harvard-Smithsonian Center for Astrophysics, 60 Garden Street, MS-72, Cambridge, MA 02138} 
\altaffiltext{4}{Astronomy Department, University of California, 601 Campbell Hall, Berkeley, CA 94720-3411}
\altaffiltext{5}{Department of Astronomy and Astrophysics, The Pennsylvania State University, 525 Davey lab, University Park, PA, 16802}

\begin{abstract}
We present a new set of photometric transformations for red stars observed with the Sloan Digital Sky Survey (SDSS) 0.5-m Photometric Telescope (PT) and the SDSS 2.5-m telescope at the Apache Point Observatory in New Mexico. Nightly PT observations of US Naval Observatory standards are used to determine extinction corrections and calibration terms for SDSS 2.5-m photometry. Systematic differences between the PT and native SDSS 2.5-m {\it ugriz} photometry require conversions between the two systems which have previously been undefined for the reddest stars. By matching $\sim 43,000$ stars observed with both the PT and SDSS 2.5-m, we extend the present relations to include low-mass stars with colors $0.6 \le r-i \le 1.7$. These corrections will allow us to place photometry of bright, low-mass trigonometric parallax stars previously observed with the PT on the 2.5-m system. We present new transformation equations and discuss applications of these data to future low-mass star studies using the SDSS.
\end{abstract}


\section{Introduction}
	
The study of low-mass stars has blossomed with the advent of large scale surveys.  Due to their low intrinsic luminosity ($\lesssim 10^{-2}L_\sun$) and the small fraction of light emitted in the optical band, most prior large area surveys have been limited to nearby objects.  Recent surveys such as the Sloan Digital Sky Survey (SDSS; \citealp{2000AJ....120.1579Y}) and the Two Micron All-Sky Survey (2MASS; \citealp{1997ilsn.proc...25S}) have now produced extensive volumes of photometry and spectroscopy on cool stars (eg. \citealp{1999ApJ...522L..61S,2002AJ....123.3409H,2004PASP..116.1105W,2004AJ....128..426W,2005PASP..117..706W,2006AJ....132.2507W,2007AJ....133..531B,c07}).  Studying these faint neighbors gives us insight into the most numerous stellar population in the Galaxy, probing both young and old subsets of low-mass stars, and placing their properties, such as abundances and dynamics, in a Galactic context \citep{2002AJ....123.3409H,2006AJ....131.2722C,met07}. Ongoing studies seek to measure the luminosity function and mass function of low-mass stars \citep{c07}, the dynamics of the thin and thick disks of the Milky Way \citep{b07}, and to characterize magnetic activity and dynamical heating in the Galactic disk \citep{2004AJ....128..426W,2006AJ....132.2507W,w07}. 

These studies are enabled by SDSS photometry due to its high sensitivity, near-infrared bandpasses, and accurate photometry which achieves a relative precision of 2-3\% \citep{2004AN....325..583I}. However, creating a suitable calibration onto an absolute system for SDSS photometry over the entire range of stellar colors poses a great challenge. US Naval Observatory (USNO) measurements of 158 standard stars form the foundation for the SDSS photometric calibration \citep{1996AJ....111.1748F,2002AJ....123.2121S}. These standards are observed on the $u^{^{\prime}} g^{^{\prime}} r^{^{\prime}} i^{^{\prime}} z^{^{\prime}}$ system and are comprised mainly of early-type ($<$ spectral class M) stars too bright to be imaged with the SDSS 2.5-m, but which have well defined magnitudes.

The SDSS Photometric Telescope (PT), with an aperture of 0.5-m, is employed to transfer the photometric calibration from the USNO measurements to the SDSS. Located alongside the SDSS 2.5-m telescope \citep{2006AJ....131.2332G} at Apache Point Observatory (APO), the PT provides nightly observations on the $u^{\prime} g^{\prime} r^{\prime} i^{\prime} z^{\prime}$ photometric system.\footnote{The PT telescope uses a $u^{\prime} g^{\prime} r^{\prime} i^{\prime} z^{\prime}$ filter set, and the PT data are first calibrated to the USNO $u^{\prime} g^{\prime} r^{\prime} i^{\prime} z^{\prime}$ system before being transformed into SDSS {\it ugriz} magnitudes.} By imaging patches of sky coincident with the SDSS nightly footprint, the PT is able to provide a robust calibration between the $u^{\prime} g^{\prime} r^{\prime} i^{\prime} z^{\prime}$ and native SDSS 2.5-m {\it ugriz} systems. \cite{2006AN....327..821T}, hereafter T06, details this three telescope calibration method, including the data reduction pipeline (MTPIPE) created for the PT.

\begin{figure}[!ht]
\centering
\includegraphics[width=4in]{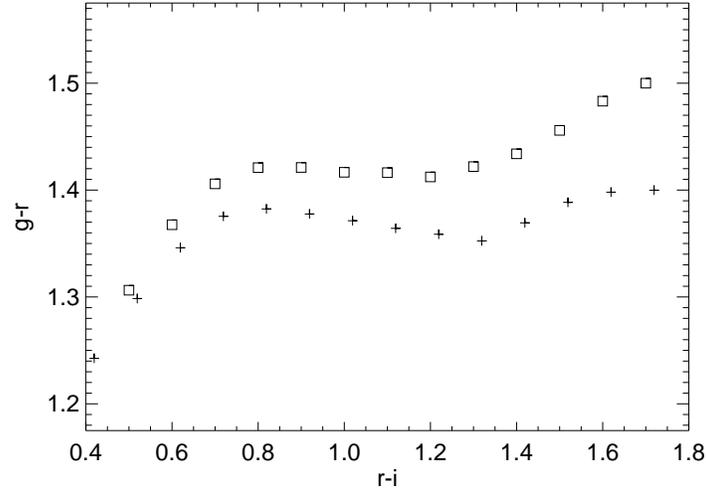}
\caption{Median $(g-r)$, $(r-i)$ colors of low-mass stars. The SDSS data shown as open squares are on the $(ugriz)_{SDSS}$ system, PT data (crosses) are on the $(ugriz)_{PT}$ system. Note the $\sim 0.1$ magnitude offset between the two stellar loci for red $(r-i)$ colors.}
\label{prob}
\end{figure}

Unfortunately, the transformations described in T06 are defined only for limited ranges in color-space. These ranges, listed in Table 1, are too blue for low-mass star studies \citep{2007AJ....133..531B,2005PASP..117..706W}. In particular, $(ugriz)_{PT} $ photometry\footnote{The expression $(ugriz)_{PT}$ denotes PT photometry which has been transformed from the native PT $u^{\prime} g^{\prime} r^{\prime} i^{\prime} z^{\prime}$ to the standard SDSS 2.5-m $ugriz$ system using the T06 calibrations. We adopt the subscript notation $(ugriz)_{SDSS}$ for native SDSS 2.5-m photometry. } diverges from the $(ugriz)_{SDSS}$ stellar locus for stars redder than $(r-i)_{SDSS} \ge 0.6$, as shown in Figure \ref{prob}. These  systematic offsets are not unexpected, due to physical differences between the $u^{\prime} g^{\prime} r^{\prime} i^{\prime} z^{\prime}$ and {\it ugriz} filters and the complex calibration between the two native telescope systems (T06). Additional correction terms are needed to rectify this systematic offset, thus placing red stars observed with the PT on the native SDSS 2.5-m system. 

An important application of these calibrations would permit PT observations of bright trigonometric parallax stars to be placed on the $(ugriz)_{SDSS}$ system. This will improve distance determinations from photometric parallaxes for low-mass stars, which would greatly benefit many of the studies described above.

In this paper we present analysis of photometry for $\sim$43,000 red point sources with both SDSS 2.5-m and PT detections. These data are used to derive transformations between the  $(ugriz)_{PT} $ and  $(ugriz)_{SDSS} $ systems for stars with $0.6 \le (r-i)\le1.7$. Our SDSS and PT sample selection is described in \S2. In \S3, we present our transformation equations for red stars between the two systems. A discussion of the implications of this improved calibration for low-mass star studies is presented in \S4.

\section{Sample Selection}
\subsection{SDSS Sample}
Our data are taken from the 8000 sq. degrees of the SDSS Data Release 5 (DR5; \citealp{AM07}). Technical descriptions of the survey can be found in the references in the introduction as well as in \cite{1998AJ....116.3040G}, \cite{2001AJ....122.2129H}, \cite{2003AJ....125.1559P}, and \cite{2002AJ....123..485S}.

To generate a photometric catalog of low-mass stars in the ($ugriz)_{SDSS}$ system,
we queried the Catalog Archive Server\footnote{http://cas.sdss.org/dr5/en/} for photometric observations classified as stars with the colors of $r - i >$ 0.5 and $i - z >$ 0.3 within the DR5 footprint. To
ensure accurate point spread function (psf) photometry, several photometric quality flags were imposed on the data.\footnote{Specifically, we required detections in BINNED1 in the $r$ , $i$ and $z$ bands.  We also selected against $r$ , $i$ and $z$ observations with the EDGE, NOPROFILE, PEAKCENTER, NOTCHECKED, PSF$\_$FLUX$\_$INTERP, SATURATED, DEBLEND$\_$NOPEAK, INTERP$\_$CENTER, COSMIC$\_$RAY or BAD$\_$COUNTS$\_$ERROR flags set, as well as observations with psf magnitude errors greater than 0.2 mag.}  The final sample contained approximately 13.6 million stars.  The brightness distribution of these objects is shown in Figure 2. Because our goal was only to compare the photometric response of red stars between the $ugriz$ and $u^{\prime} g^{\prime} r^{\prime} i^{\prime} z^{\prime}$ systems, no corrections for galactic extinction were used.

\subsection{PT Sample}
During normal photometric operations on the SDSS 2.5-m telescope, the PT automatically images overlapping patches of the sky for calibration. These patches are roughly $15^\circ $ apart and cover every stripe (great circle path of scan) in the SDSS survey. Typically these images are used to create photometric zero-point `anchors' for the SDSS data onto the USNO system. However, these patches also contain many stars  which are not used in the SDSS calibration, but are reduced and saved by MTPIPE (T06). Over five million background star observations were made with the PT during the first six years of SDSS observations. These reduced MTPIPE data are available via the SDSS Data Archive Server.\footnote{http://das.sdss.org/PT/} 

After removing stars which MTPIPE flagged as having bad photometry (magnitude $= -100$), the PT sample was selected using two color cuts. Low-mass candidates were selected by requiring $(r-i)_{PT} > 0.5$ \citep{2007AJ....133..531B}. A further color cut of $1.1 \le (g-r)_{PT} \le 1.7$ was used to isolate the low-mass stellar locus, removing carbon stars \citep{1999AJ....117.2528F}. This process yielded $\sim 345,000$ stars with red colors which have accurate PT magnitudes. The PT sub-sample has mean magnitude errors of $\sigma(g_{PT})= 0.05$, $\sigma(r_{PT}) = 0.07$, $\sigma(i_{PT}) = 0.08$, and $\sigma(z_{PT}) = 0.06$. We did not use the $u_{PT}$ data due to the lack of good photometry for these red stars; only $\sim$12,000 stars in the PT sample have good $u_{PT}$ photometry ($\sigma(u_{PT}) \le 0.05$).

\subsection{Matched Sample}
\begin{figure}[!ht]
\centering
\includegraphics[width=6in]{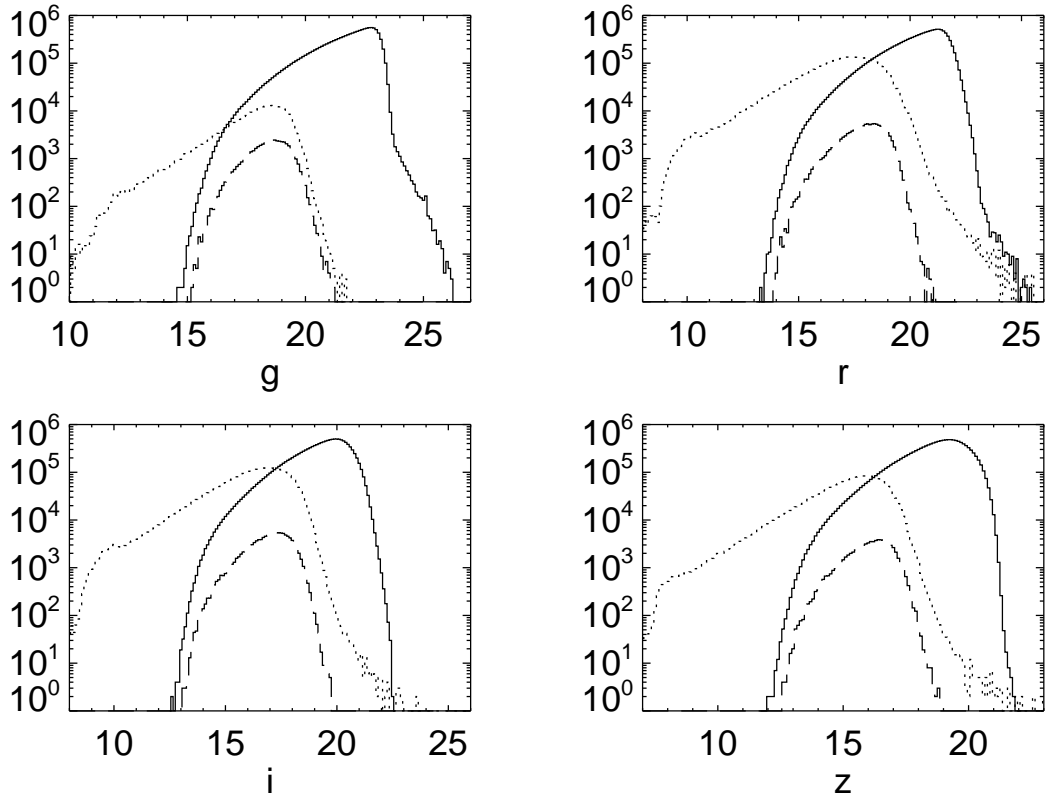}
%\centerline{	
%	\mbox{\includegraphics[width=3in]{hist_g}}
%	\mbox{\includegraphics[width=3in]{hist_r}} }
%\centerline{
%	\mbox{\includegraphics[width=3in]{hist_i}}
%	\mbox{\includegraphics[width=3in]{hist_z}} }
\caption{Histograms of the full five million PT (dotted line) and 13.6 million SDSS (solid line) data sets. Note the brighter limits of the smaller PT telescope. The matched sub-sample of $\sim$43,000 stars is shown as the dashed line for each filter. }
\label{hists}
\end{figure}

To facilitate matching between the two samples, we imposed a bright magnitude limit on the PT sample to isolate stars potentially observed by both telescopes (See Figure \ref{hists}). In particular, the following magnitude cuts were used on the PT sample: $15.4 \le g_{PT} \le 21.5$ ; $14.6 \le r_{PT} \le 20.0$ ; $14.0 \le i_{PT} \le 19.7$ ;  $14.1 \le z_{PT} \le 19.5$. 

The PT and SDSS samples were matched by celestial position, yielding $42,912$ PT-2.5-m matches within a search radius of $0.5''$. The SDSS 2.5-m astrometry is accurate to 0.045", while the PT astrometry is accurate to  $\sim$1".  As we are not trying to find a complete sample, we chose a conservative search radius (0.5") to minimize mismatches, and still obtain a large number of matches. Since typical proper motions of these stars were small over the limited time range of PT observations ($2002-2007$), the loss of some high proper-motion stars was negligible.

\section{Results}
\subsection{Determining Offsets}

\begin{figure}[!ht]
\centering
\includegraphics[width=6in]{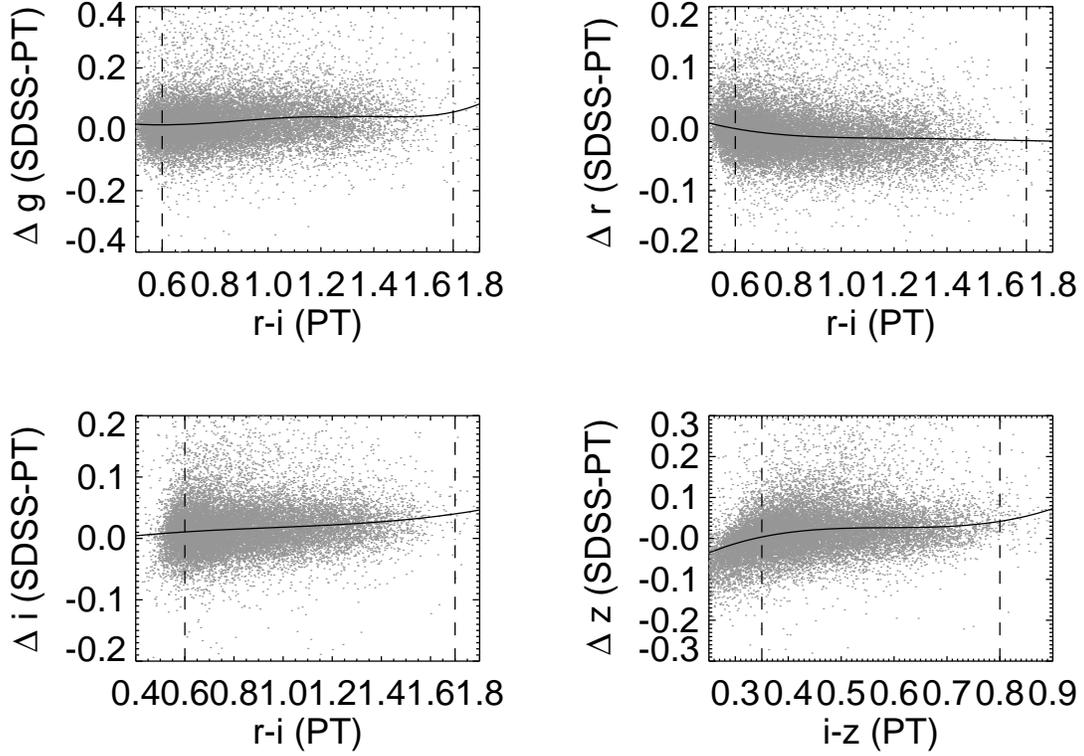}
%\centerline{
%	\mbox{\includegraphics[width=3in]{final_delta_g}}
%	\mbox{\includegraphics[width=3in]{final_delta_r}} }
%\centerline{
%	\mbox{\includegraphics[width=3in]{final_delta_i}}
%	\mbox{\includegraphics[width=3in]{final_delta_z}} }
\caption{Difference in PT and SDSS magnitudes as a function of $(r-i)_{PT}$ and $(i-z)_{PT}$ color. Error weighted polynomial least-squares fits are shown as solid lines in each panel. The vertical dashed lines indicate the color ranges over which the fits are reliable.}
\label{slopes}
\end{figure}

Using the matched sample described in \S2.3, the magnitude differences between the two systems were determined for each $griz$ filter as functions of PT color (see Figure \ref{slopes}).  Least-squares polynomial fits to the data provide quantitative offsets between the $(griz)_{PT} $ and  $(griz)_{SDSS}$ systems (see \S3.2). Two-part piecewise functions were used for the $g_{PT}$ and $r_{PT}$ corrections which were functions of $(r-i)_{PT}$. These fits both had breaks at $(r-i)_{PT} = 1.25$ and were found to be better fits to the data than single polynomials.

Systematic selection of stars in the blue end of the data shown in Figure \ref{slopes} occurs due to the color cuts chosen in the PT and SDSS sample selection. Our fitting regions, shown in Figure 3 and listed in Table 2, avoid these areas. To remove any strong outliers, (i.e. visual binaries, mismatches, and flares), our fits are weighted by photometric error and are based only on data with a photometric error less than $0.05$ magnitudes in every color band for each observation.

\subsection{Transformations}
The following equations produce the fits shown in Figure \ref{slopes}. These fits provide the best transformation of $(griz)_{PT}$ data to the native $(griz)_{SDSS}$ system for red stars. 

\noindent
% \begin{align}  %g
\begin{eqnarray}
  g_{SDSS}(0.6 \le (r-i)_{PT} < 1.25) = & g_{PT}+0.142-0.514(r-i)_{PT}+0.647(r-i)_{PT}^2\nonumber \\ 
  & -0.241(r-i)_{PT}^3 \pm0.072,\nonumber \\
  g_{SDSS}(1.25 \le (r-i)_{PT} \le 1.7) = & g_{PT}-2.511+5.391(r-i)_{PT}-3.787(r-i)_{PT}^2\nonumber \\
  & +0.885(r-i)_{PT}^3 \pm0.082
%\end{align}
\end{eqnarray}
%\begin{align} %r
\begin{eqnarray}
  r_{SDSS}(0.6 \le (r-i)_{PT} < 1.25) = & r_{PT}+0.116-0.339(r-i)_{PT}+0.301(r-i)_{PT}^2\nonumber \\
  & -0.091(r-i)_{PT}^3\pm0.043,\nonumber \\
  r_{SDSS}(1.25 \le (r-i)_{PT} \le 1.7) = & r_{PT}+0.003-0.009(r-i)_{PT} \pm0.055 
%\end{align}
\end{eqnarray}
%\begin{align} %i
\begin{eqnarray}
 i_{SDSS}(0.6 \le (r-i)_{PT} \le 1.7) = & i_{PT}-0.019+0.082(r-i)_{PT}-0.068(r-i)_{PT}^2\nonumber \\
 &+0.024(r-i)_{PT}^3 \pm 0.033
%\end{align}
\end{eqnarray}
%\begin{align} %z
\begin{eqnarray}
 z_{SDSS}(0.35 \le (i-z) \le 0.8) = & z_{PT}-0.285+1.566(i-z)_{PT}-2.637(i-z)_{PT}^2\nonumber \\
 & +1.488(i-z)_{PT}^3\pm0.050
%\end{align}
\end{eqnarray}

These equations are valid for the color ranges listed in Table 2, and should {\bf only} be employed within these ranges. Also, these corrections are applicable only to PT observations which have been reduced by MTPIPE and not USNO $u^{\prime} g^{\prime} r^{\prime} i^{\prime} z^{\prime}$ data. After applying the transformations, there is a significant improvement in the average colors of the stellar locus of red stars. As shown in Figure 4, the PT data from Figure 1 align with the SDSS stellar locus when the new transformations are employed. Although the corrections from Equations 1-4 are small, the offsets amount to more than 0.1 mag in $(g-r)$ for the reddest stars. Corrections in the $(r-i)$, $(i-z)$ diagram (Figure 5) are less substantial, but still significant.

\section{Discussion}
We have matched a large sample of background red stars observed with the PT to photometric observations of the same stars in the SDSS DR5 catalog. We present transformations based on magnitude differences between the $(griz)_{PT}$ and $(griz)_{SDSS}$ photometric systems as functions of color.

\begin{figure}[!ht]
\centering
\includegraphics[width=4in]{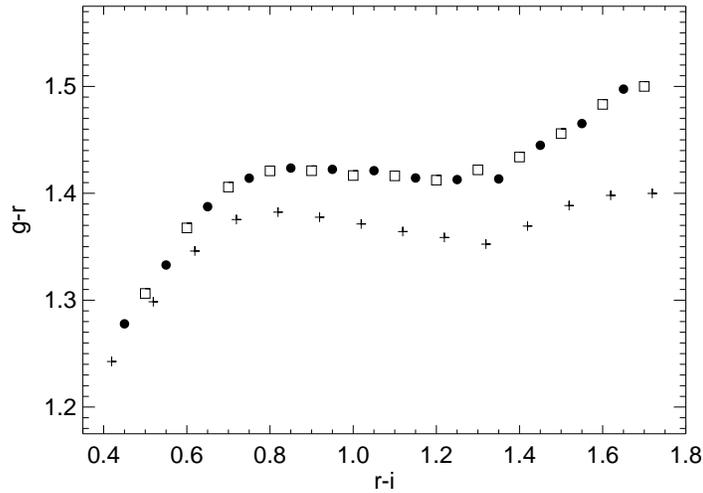}
\caption{Median $(g-r)$, $(r-i)$ colors of low-mass stars as in Figure \ref{prob}. $gri_{SDSS}$ data are shown as open squares. $gri_{PT}$ data prior to our corrections are shown as crosses, while $gri_{PT}$ data transformed using the equations in \S3.2 are shown as filled circles.}
\label{money1}
\end{figure}

\begin{figure}[!ht]
\centering
\includegraphics[width=4in]{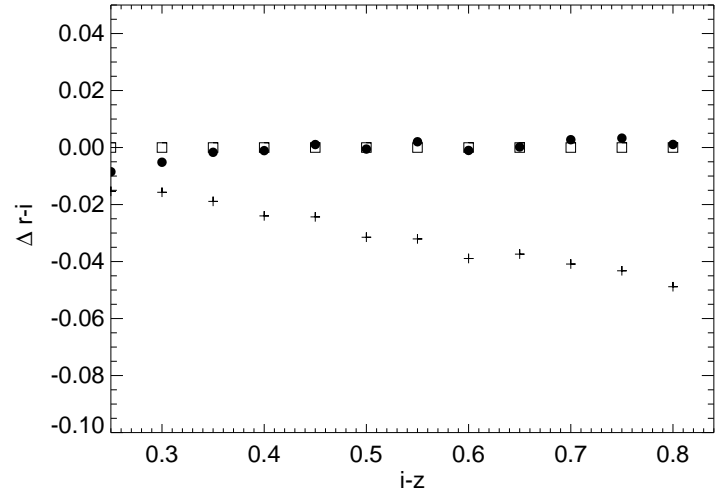}
\caption{Median $\Delta(r-i)$, $(i-z)$ residuals of low-mass stars. SDSS - SDSS data are shown as open squares. SDSS - PT data on the SDSS system prior to our corrections are shown as crosses, while SDSS - PT data transformed using the equations in \S3.2 are shown as filled circles. }
\label{money2}
\end{figure}

These transformations provide a significant improvement in the photometric calibration of PT data onto the SDSS 2.5-m system for red stars. Our results augment the initial transformations described in T06 which have already been applied to all publicly available PT data. The systematic color offsets for red stars, corresponding to spectral types M0 through M5 \citep{2007AJ....133..531B}, which have previously been seen between the PT and SDSS (see Figures \ref{money1} and \ref{money2}) are corrected when our transformations are applied.

The ability to place the $(griz)_{PT}$ photometry onto the native SDSS 2.5-m system allows PT observations to serve as a `bright extension' of the SDSS survey.  In particular, PT observations of red stars with measured trigonometric parallaxes can now be reliably transformed onto the SDSS 2.5m system with a typical accuracy $<10\%$, allowing an improved definition of the color-magnitude relation for low-mass stars in native SDSS color-space \citep{2002AJ....123.3409H,2005PASP..117..706W,2006PASP..118.1679D}. We expect these results to be useful in analyzing targeted PT observations of red dwarf parallax standards \citep{2002AAS...20111511W, gol07}, as well as serendipitous PT observations of stars with Hipparcos parallaxes.

\acknowledgments
The authors would like to thank Douglas Tucker and David Golimowski for their insightful comments, suggestions, and assistance with MTPIPE. The authors acknowledge the support of NSF grants AST02-05875, AST06-07644, and NASA ADP grant NAG5-1211. K.R.C. acknowledges support for this work was provided by NASA through the Spitzer Space Telescope Fellowship Program, through a contract issued by the Jet Propulsion Laboratory, California Institute of Technology under a contract with NASA. A.A.W. acknowledges the support of NSF grant AST05-40567.

This project made extensive use of SDSS data. Funding for the SDSS and SDSS-II has been provided by the Alfred P. Sloan Foundation, the Participating Institutions, the National Science Foundation, the U.S. Department of Energy, the National
Aeronautics and Space Administration, the Japanese Monbukagakusho, the
Max Planck Society, and the Higher Education Funding Council for
England. The SDSS Web Site is http://www.sdss.org/.

The SDSS is managed by the Astrophysical Research Consortium for the
Participating Institutions. The Participating Institutions are the
American Museum of Natural History, Astrophysical Institute Potsdam,
University of Basel, Cambridge University, Case Western Reserve
University, University of Chicago, Drexel University, Fermilab, the
Institute for Advanced Study, the Japan Participation Group, Johns
Hopkins University, the Joint Institute for Nuclear Astrophysics, the
Kavli Institute for Particle Astrophysics and Cosmology, the Korean
Scientist Group, the Chinese Academy of Sciences (LAMOST), Los Alamos
National Laboratory, the Max-Planck-Institute for Astronomy (MPIA),
the Max-Planck-Institute for Astrophysics (MPA), New Mexico State
University, Ohio State University, University of Pittsburgh,
University of Portsmouth, Princeton University, the United States
Naval Observatory, and the University of Washington.

We finally would like to thank the anonymous referee for their helpful comments and questions.

\clearpage

%figures

\clearpage

%tables
\begin{deluxetable}{c}
\tablecolumns{1}
\tablewidth{0pt}
\tablecaption{Tucker et al. (2006) PT Standard Star Color Ranges}
\tablehead{
	\colhead{Color}}
\startdata
$0.70 \le (u - g)_{PT} \le 2.70$\\
$0.15 \le (g - r)_{PT} \le 1.20$\\
$-0.10 \le (r - i)_{PT} \le 0.60$\\

$-0.20 \le (i - z)_{PT} \le 0.40$
\enddata
\end{deluxetable}

\begin{deluxetable}{c}
\tablecolumns{1}
\tablewidth{0pt}
\tablecaption{Applicable Color Ranges for Equations 1-4}
\tablehead{
	\colhead{Color}}
\startdata
$1.20 \le (g - r)_{PT} \le 1.55$\\
$0.60 \le (r - i)_{PT} \le 1.70$\\
$0.35 \le (i - z)_{PT} \le 0.80$
\enddata
\end{deluxetable}

\end{document}